\documentclass[runningheads]{llncs}
\usepackage[T1]{fontenc}

\usepackage{graphicx}
\usepackage{enumitem}

\begin{document}

\title{Usage Control in Public Data Exchange Infrastructures for Future Digital Governance}

\titlerunning{Usage Control for Next-Gen Digital Governance}

\author{
Oleksandr Kosenkov\inst{1} \and
Fathiyeh Faghih\inst{2} \and
Eric Jackson\inst{3} \and
Theo Dimitrakos\inst{4}
}

\authorrunning{O. Kosenkov et al.}

\institute{
Blekinge Institute of Technology, Karlskrona, Sweden\\
\email{oleksandr.kosenkov@bth.se}
\and
Birmingham City University, Birmingham, United Kingdom\\
\email{fathiyeh.faghih@bcu.ac.uk}
\and
Tallinn University of Technology, Tallinn, Estonia\\
\email{eric.jackson@taltech.ee}
\and
University of Kent, Canterbury, United Kingdom\\
\email{t.dimitrakos@kent.ac.uk}
}

\maketitle

\begin{abstract}
Public data exchange infrastructures, such as X-Road, are essential to data-intensive digital public services and the EU digital economy. Their next generation requires advanced data usage control (DUC) mechanisms that address emerging requirements, including privacy by design, to mitigate privacy threats inherent in large-scale automated data exchange. Although research within the individual disciplines already addresses many of the emerging requirements, significant practical challenges arise at the intersection of the disciplines.

This paper analytically examines the practical challenges and requirements of data usage control in public data exchange infrastructures. We examine the Estonian X-Road system and emerging use cases through an interdisciplinary lens that integrates formal verification and monitoring of usage control, requirements engineering, and digital governance research. We formulate key requirements and propose a conceptual vision for developing usage control in public data-exchange infrastructures to enable future digital-governance systems to comply with privacy requirements by design.

\end{abstract}

\section{Introduction}
\label{sec:intro}
Public and private organizations rely heavily on data in the development and delivery of products and services. In the public sector, effective use of data is fundamental to digital service delivery and increasingly important for adopting AI and other data-intensive technologies. However, data is typically distributed across multiple government organizations that traditionally operate in silos, making secure and efficient data sharing challenging.

The adoption of data-intensive service delivery models also increases privacy risks. Under the General Data Protection Regulation (GDPR), organizations must ensure both the security of personal data and individuals' control over their data. As digital public services become increasingly automated, balancing data accessibility with privacy protection becomes essential.

Countries with highly digitalized public sectors have implemented public data exchange infrastructures such as X-Road in Estonia, while initiatives such as European data spaces aim to enable interoperable and sovereign data sharing without centralized storage. However, many existing infrastructures were designed for relatively static service delivery and provide limited support for increasingly automated, AI-enabled public services.

Advanced DUC can contribute to emerging digital government by supporting context-sensitive control over data access and subsequent use, thereby supporting privacy by design. However, existing DUC research has only partially addressed privacy by design and the needs of next-generation data-exchange infrastructures. In this paper, we investigate, from an interdisciplinary perspective, the requirements for operationalizing DUC in future public data-exchange infrastructures and propose a corresponding vision for its application. Our analysis is informed by three complementary research perspectives: formal verification and monitoring of DUC, requirements engineering, and digital governance. We use Estonia's X-Road as an illustrative and analytically useful example of a public data-exchange infrastructure because of its maturity and international adoption. We used a problem-centered design-science approach to analytically and iteratively perform the following research activities, which produced the contributions reported in this paper:
\begin{itemize}[nosep]
    \item[(i)] reviewed the state of the art and state of DUC implementation in X-Road (Section~\ref{sec:background});
    \item[(ii)] identified emerging digital-governance scenarios exposing limitations of current DUC models from an interdisciplinary perspective (Section~\ref{sec:cases});
    \item[(iii)] derived an initial set of cross-disciplinary requirements for bridging the identified gaps (Section~\ref{sec:requirements});
    \item[(iv)] developed a conceptual vision for a closed-loop architecture that integrates policy synthesis, verification, enforcement, and monitoring to support emerging use cases and requirements including privacy by design (Section~\ref{sec:vision}).
\end{itemize}

\section{Background}\label{sec:background}
\paragraph{Data spaces and interoperability}
The development of data-exchange infrastructures is an important component of the EU’s digital and data strategy, with data spaces—federated ecosystems for sovereign data sharing—serving as a key technological enabler~\cite{reiberg2022dataspace}. European data spaces aim to enable secure and interoperable data sharing across public and private sectors while preserving data sovereignty~\cite{ec2020european,ec2025data}. In data spaces DUC is based on technologies such as ODRL-based policies~\cite{fraunhofer2026mydata}. While these approaches provide important foundations for controlled data sharing, they primarily target predefined sharing scenarios and have limited support for dynamic public-sector environments.
According to the European Interoperability Framework, interoperability--the ability of different stakeholders’ information systems to exchange data--should be implemented at the technical, semantic, organizational, and legal levels~\cite{idsa2025}. 

\begin{figure}
    \centering
    \includegraphics[width=.75\linewidth]{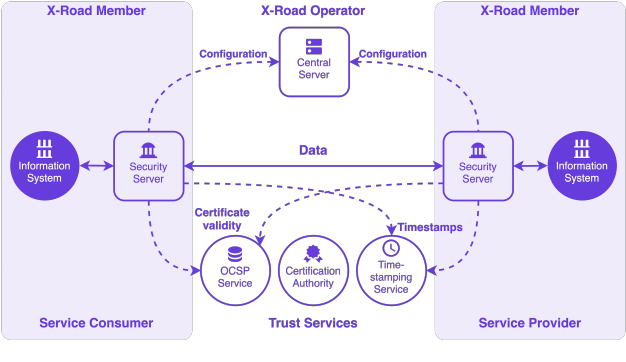}
    \caption{Schematic depiction of data exchange in X-Road including security servers implementing access control.}
    \label{fig:xroadExchange}
\end{figure}

\paragraph{Public data exchange infrastructures for digital governance}
X-Road is an operational public data exchange infrastructure playing a key role in the digitalization of the public sector and delivery of public services in Estonia.
X-Road is a centrally managed data exchange layer between the distributed information systems of governmental agencies. Its main purpose is to enable public authorities to deliver digital services to citizens and businesses by supporting government-to-government and government-to-citizen and government-to-business exchange of data. For this, X-Road ensures decentralized peer-to-peer communication between databases and service endpoints maintained by different public authorities (see Figure~\ref{fig:xroadExchange} for an overview). The first version of X-Road was launched in 2001 by the Information System Authority of Estonia~\cite{kalja2002x,blake2021historical}, and since then, it has been adopted by about 25 countries worldwide mainly for the implementation of digital public services, but also for private sector use cases~\cite{xroad2020planetcross}. 

Initially, access control in X-Road required data-exchange agreements specifying what data would be shared, when, and by whom, and was technically implemented through the Security Server’s configuration database~\cite{puura2025advancing}. Access policies have been defined using client identifiers, service codes, HTTP verbs, and request paths~\cite{xroad2024making}. In 2024, Open Digital Rights Language (ODRL) was additionally implemented for access policies specification, and further migration to ODRL was planned~\cite{xroad2024making}. However, the overall approach to data access policies remains organization-focused; in particular, access policies are primarily static and related to the predefined data required for the delivery of specific public services and agreements between the public authorities. 

\paragraph{Privacy by Design}
Privacy by design and principle of accountability, mandated by GDPR, require demonstrable implementation of technical privacy safeguards in system design. Consequently, public data exchange infrastructures require transparent and verifiable privacy mechanisms rather than opaque processing. Despite the importance of this legal requirement, proposed methods for implementing privacy by design remain largely ad hoc and non-systematic; however, some approaches emphasize the need to involve legal experts~\cite{kosenkov2025privacy}.

\section{Emerging cases of next-generation digital governance}\label{sec:cases}
Next, we describe the emerging cases for public data exchange infrastructure and challenges they create for data usage control.

\paragraph{Proactive public digital governance and services}
Traditional public service delivery is largely reactive. Government-to-citizen/ business digital services typically require the user to initiate the interaction and invest effort. A next generation of digital public services aims to be proactive, leveraging interoperability across organizational data silos so that the public sector can reuse data it already holds across ministries and agencies and reduce or eliminate unnecessary steps. In such proactive designs, the occurrence of a qualifying life event (e.g., child birth, marriage) serves as a trigger for automated or semi-automated service delivery, shifting the burden from users to the public sector while preserving appropriate safeguards for consent, legality, and accountability.

However, proactive public services form part of a broader strategy for proactive government that anticipates, rather than merely reacts to, societal and economic developments and individual citizens’ needs~\cite{mcbride2023proactive}. This also implies personalization of data exchange flows of digital services depending on individuals' needs. One technical implication of such a strategy is that the data exchange workflows supporting next-generation proactive services cannot be fully specified in advance. Unlike traditional digital services, where interactions between information systems follow predefined processes, proactive services require data to be exchanged across organizations depending on the context of a particular life event and the individual circumstances of the service recipient. As a result, the participating systems, required datasets, and sequence of interactions may differ from case to case. This means that data exchange relationships are established dynamically rather than through fixed integrations. In such environments, ensuring that data is processed appropriately becomes significantly more complex. Mechanisms are therefore required that can automatically enforce data access and usage constraints across organizational boundaries, taking into account legal mandates, institutional responsibilities, and user consent.

\paragraph{Public-private data exchange}
Future public data exchange infrastructures are expected to involve stronger collaboration between public and private organizations. While inter-organizational agreements and traditional access control may be sufficient in relatively small ecosystems, broader participation requires fine-grained, real-time authorization and mechanisms that ensure data is subsequently used only for its intended purposes.

\paragraph{Flexible AI agent-based governance}
Estonia and other countries using X-Road-based systems, such as Ukraine, are already developing strategies for an ``agentic state''—i.e., a model in which AI systems proactively coordinate administrative tasks, help interpret and apply policy constraints, and scale service delivery while humans retain accountability for discretionary judgment and oversight. In this view, agentic systems do not merely provide predefined public services; they increasingly become intermediaries that compose service journeys, route requests across agencies, and trigger transactions spanning multiple registries and information systems—raising a governance challenge that is as much about control and accountability as it is about automation.

A representative example is Bürokratt, an Estonian initiative towards agentic government. While current implementations mainly provide information support, the long-term vision is to enable AI agents to coordinate interactions across multiple public services and organizations. Such evolution fundamentally changes how access and usage decisions must be enforced.
Although X-Road provides secure data exchange and citizen-facing transparency mechanisms such as Data Tracker, these mechanisms primarily address secure access rather than controlling how data is subsequently used in AI-driven workflows.

In an agentic public sector scenario, the “decision point” for compliance may shift from a single system boundary to an end-to-end service journey dynamically composed by agents. As a result, future agentic assistants that leverage X-Road-enabled registries will require DUC mechanisms that can express and enforce purpose limitation, consent constraints, delegation scopes, and auditability across multiple systems—supporting privacy by design not only at the level of infrastructure security, but also at the level of how data is used throughout an automated, cross-organizational workflow.

Effective data usage control for agentic AI requires moving beyond one-time access decisions towards a continuous enforcement model based on mutable attributes, ongoing obligation tracking, and real-time policy evaluation. Such a model could support the data minimization principle through Minimum Viable Data, whereby agents are granted access only to the specific, contextually relevant data required for a given task, rather than broad access to entire databases. 

\paragraph{Privacy by design}
While existing access control implementation in X-Road provides rich opportunities for data exchange and basic trustworthiness, it is not sufficient to fully operationalize the demands of privacy by design and make privacy an integral property of the data exchange.

To illustrate these challenges, we examine a recent data-access incident in Estonia. In 2025, it was reported that thousands of requests for banking data were made through the Estonian public data exchange infrastructure without clear legal proceedings in place~\cite{erree2025justice}, and about two thousand of those requests were made without legal basis~\cite{erree2025justice2}. As a result, many entrepreneurs needed to request information about access to their data through the court as the access procedure was not clearly regulated~\cite{erree2025entrepreneurs}. Banks participating in the automated data exchange claimed approximately €247 million in damages resulting from unlawful access to their customers’ data~\cite{erree2025lhv}. One reported explanation was that the Financial Intelligence Unit, which requested the data without a clear legal basis, had originally been authorized to submit such requests while operating under one governmental agency. After it was transferred to another agency, it lost the legal authority to do so but retained the technical capability. Later, it was announced that such data queries will be logged and available for review. It was also suggested that even when request is executed according to existing proceedings, there should be sufficient opportunity to check the legality of each specific request~\cite{err2025ministry}. However, the Ministry of Justice of Estonia disagreed, stating that such legality could be checked by courts if required, but cannot be checked by the ministry executing requests and providing the automated data access.

While Estonian public data exchange infrastructure provides visibility into the sharing of the personal data~\cite{erree2025state} and post-fact identification of data breaches, it does not block non-compliant queries and proactively prevent data breaches. Moreover, such a reactive mechanism has a limited usability requiring control from users. We suggest that for data exchange infrastructure to be completely compliant with the privacy by design principle, it is necessary to ensure granular data access and automated prevention of non-compliant access.

\section{DUC requirements for next-generation data exchange cases}\label{sec:requirements}
The scenarios discussed in Section~\ref{sec:cases} show that traditional access control is insufficient for next-generation public data exchange infrastructures. DUC technologies must therefore satisfy a broader set of requirements supporting dynamic, cross-organizational, and AI-mediated data sharing. Based on these scenarios, we derive the following requirements.

\paragraph{R1: Normative Alignment with Regulatory and Organizational Constraints}
Public data exchange infrastructures operate in environments with different regulations and contractual norms potentially applicable to different infrastructure participants~\cite{talmoudi2024compliance}. Legal norms are usually intentionally abstract, are goal-oriented, contain high-level principles (e.g.,  data minimization), and do not prescribe fixed technical solutions.  Simultaneously, DUC enforcement mechanisms operate at the level of concrete system behavior, which requires formalization of regulatory norms. Still, such formalization of regulatory provisions is considered complex even in specialized research areas such as AI\&Law and legal informatics. In these areas, it is recognized that there is no one single correct interpretation of regulations and for this reason formalization methods should support multiple interpretations, be transparent and comprehensible~\cite{novotna2022evaluation}. 

DUC must therefore enable alignment between abstract regulatory norms, organizational policies on the one hand, and concrete usage policies and decisions on the other hand. It should be possible to demonstrate that enforcement mechanisms are grounded in identifiable legal and organizational requirements. Accordingly, DUC policy synthesis should be aligned with the legal interpretation process and take into account the legal knowledge required to go from abstract norms to concrete case-specific policies and decisions~\cite{kosenkov2021vision}.
Existing regulatory requirements engineering approaches provide useful foundations but do not fully support traceable synthesis of executable usage-control policies.

\paragraph{R2: Context-Sensitive and Fine-Grained Usage Decisions}
Some access and usage control models already account for data processing context to a certain degree. However, privacy by design and next-generation public data infrastructures create a greater need for context awareness.

Legal interpretation for privacy by design heavily relies on the context of data processing. A data processor may, in practice, qualify as a joint controller if it effectively determines the purposes and means of processing together with another party. Such determinations cannot be made in the abstract; they require the availability of extensive contextual attributes.

Another challenge in the public data exchange infrastructure is the involvement of different organizations (e.g., the X-Road ecosystem included approximately 600 organizations in 2021~\cite{blake2021historical}). Accordingly, the application of role-based or attribute-based control becomes challenging due to the compatibility of roles and attributes across different organizations. Moreover, in some cases, the application of new attributes can be required specifically for data exchange purposes. In an example of data access by a law enforcement agency provided above, an additional attribute would be required to identify if a specific officer is investigating a case related to the person whose data is requested. Otherwise, any of the investigators could get access to the data without a sufficient basis.

Moreover, integrating agentic AI with data-sharing infrastructures introduces additional layers of complexity. These include increased context sensitivity, richer and use-case-specific system states, stricter purpose limitations, adaptive permission management, and the need to maintain data provenance throughout data usage and AI inference processes. These additional layers increase the capability gap between the limited use of ODRL and Rego in data spaces and the emerging use case requirements.

 The usage control system must support context-aware decision-making based on dynamically evolving attributes and must formally account for attribute mutability over time. This entails:
\begin{itemize}[nosep]
    \item Modeling contextual attributes (e.g., case identifiers, legal basis, purpose limitation, delegation scope).
    \item Ensuring that attribute changes are semantically consistent with policy constraints.
    \item Supporting continuous re-evaluation of policies as contextual attributes evolve.
\end{itemize}

\paragraph{R3: Lifecycle and Continuity of Control}

In proactive and AI-mediated governance environments, data usage cannot be treated as a single, one-time access event. Instead, data may flow across multiple systems, organizations, and time periods as part of a larger service process. Access to data can trigger additional obligations, conditions may need to be re-evaluated during execution, and further processing may introduce new compliance requirements. In agentic AI settings, where multiple agents collaborate to perform tasks, authorization must therefore remain valid throughout the entire lifecycle of task execution. This requires continuous monitoring and enforcement of delegated permissions rather than a single authorization decision at the moment of access.

DUC mechanisms must therefore support continuity of control throughout the lifecycle of data usage. Control should extend beyond the initial authorization decision to cover ongoing and post-access phases. This includes the ability to account for evolving contextual conditions and responsibilities that arise during or after data processing. Existing policy languages such as ODRL and Rego primarily support point-in-time authorization and provide limited support for continuous lifecycle enforcement, dynamic obligations, and post-access control required by next-generation public data exchange infrastructures.

\paragraph{R4: Assurance and Compliance Guarantees}
Given the complexity of distributed infrastructures, it cannot be assumed that policy enforcement will always operate as intended. Misconfiguration, organizational restructuring, integration errors, or incomplete modeling of regulatory requirements may result in unintended permissions or unlawful data processing.

Usage control technologies must therefore provide assurance capabilities that enable stakeholders to detect inconsistencies, unintended authorizations, and non-compliant behavior. Beyond mere enforcement, the infrastructure should enable justified confidence that declared requirements are correctly reflected in operational practice. Assurance is essential for maintaining trust in public-sector digital infrastructures.

\paragraph{R5: Adaptability to Organizational and Regulatory Evolution}
Public data exchange infrastructures operate in dynamic environments.  DUC technologies must therefore support adaptability. It should be possible to revise and refine usage policies and constraints without undermining consistency, compliance, or institutional trust. For example, if regulatory changes occur, traceability between regulatory norms and policies is instrumental to reassess policies and verify compliance implementation.

\paragraph{R6: Transparency and Explainability}
Despite automation, regulatory compliance ultimately requires human accountability. Legal interpretation, risk management decisions (e.g., selection of compensatory privacy controls, assessment of residual privacy risks), and audit processes necessitate transparency. Consequently, the usage control framework must provide transparency and explainability mechanisms that enable the relevant stakeholders (including legal experts, compliance officers, and auditors) to understand, validate, and approve policy synthesis or review policies and usage control decisions retrospectively. This entails:

\begin{itemize}[nosep]
    \item Human-readable policy representations,
    \item Traceability links from norms to rules,
    \item Explanations of static and runtime analysis results,
    \item Controlled change management for policy updates.

\end{itemize}

Transparency is particularly important when resolving conflicts between regulatory requirements, organizational goals, and automated decision-making processes. Transparency and explainability of the resulting policies should enable stakeholders to review, verify, and approve any subsequent changes.

Traceability is also instrumental for explainability, as a single regulatory norm can be traced to multiple policy statements. 

\paragraph{R7: Enabling Human-in-the-Loop Oversight}
The need to align policy synthesis with legal interpretation and knowledge (R1), together with the need for context awareness (R2), motivates the development of human-in-the-loop mechanisms. Other aspects of DUC also require human involvement. Even where parts of the interpretation process can be partially automated, human oversight remains necessary to ensure that the resulting rules align with the organization’s legal risk strategy and that compliance is considered consistently within the organization. 

In practice, privacy and compliance considerations are often introduced as an afterthought. Resolving conflicts between policies derived from business requirements and those derived from privacy-compliance requirements demands informed human judgment rather than purely formal reasoning. Moreover, GDPR compliance requires a combination of technical and organizational measures. Therefore, technical usage-control mechanisms must be documented and explained in a way that allows them to be integrated with complementary organizational safeguards.
For these reasons, enabling human oversight in data usage control is not merely a design preference but a structural requirement for compliance-oriented systems.

\section{Vision: Closed-Loop DUC in Public Data Exchange}\label{sec:vision}

The cases discussed in Section 3 and the requirements derived in Section 4 point toward a common conclusion: privacy-by-design in public data exchange cannot be achieved through isolated enforcement components. What is needed instead is a continuous and accountable control cycle that connects regulatory interpretation, policy synthesis, formal verification, runtime monitoring, and systematic refinement.
\begin{figure}
\centering
\includegraphics[width=\textwidth]{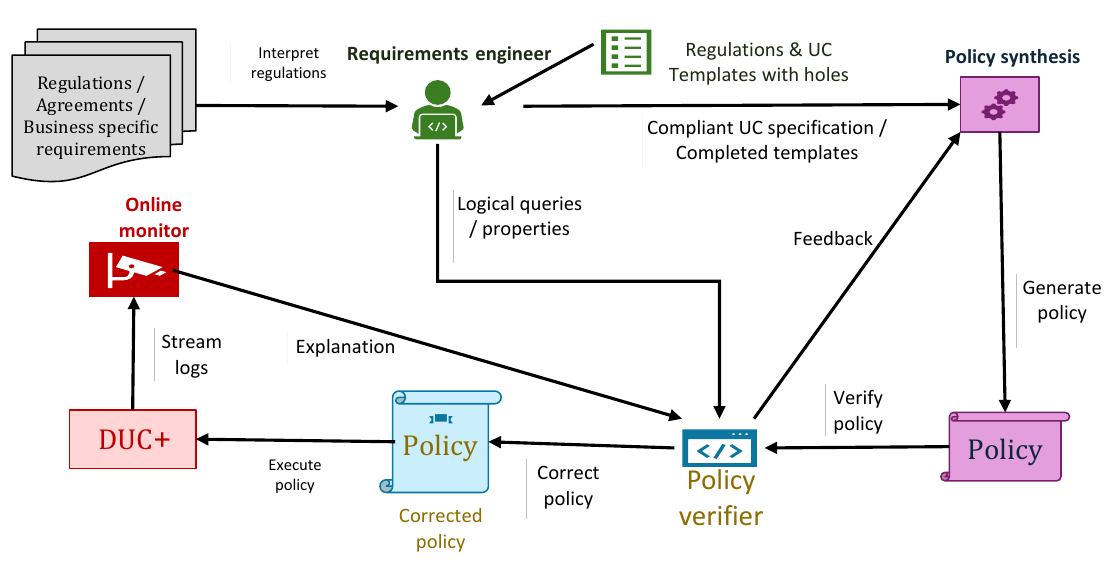}
\caption{Schematic depiction of the proposed vision for a closed-loop data usage control.}
\label{fig:arch}
\end{figure}
We therefore propose a closed-loop usage control framework tailored to public data exchange infrastructures. Rather than treating compliance as a one-time configuration task, the framework operationalizes it as an iterative governance process.  The framework consists of four interconnected layers linked by structured feedback mechanisms:

\paragraph{Regulatory-to-Policy Synthesis Layer}
    
(Addresses R1, R5, R6)

The first layer operationalizes normative alignment (R1) by providing a structured process for transforming abstract regulatory norms and data-sharing agreements into executable usage control policies. Rather than treating policy generation as a purely technical task, the synthesis process makes explicit the interpretation steps required to bridge legal language and formal enforcement constructs. The process is intentionally human-supervised: a requirements engineer  interprets regulatory requirements and completes structured templates before policy synthesis. This layer supports:
\begin{itemize}[nosep]
    \item Structured regulatory templates that capture compliance-relevant constraints,
\item Policy templates representing enforceable usage control constructs,
\item A synthesis mechanism that derives concrete rules from structured requirements,
\item Explicit traceability links from norms to policy statements.
\end{itemize}

Traceability and structured policy synthesis support transparency, adaptability, and accountable human oversight throughout policy evolution.
\paragraph{ Policy Assurance and Verification Layer}

(Addresses R1, R4, R6)

The second layer provides formal assurance before deployment. While the synthesis layer establishes traceable policy derivation, assurance ensures that synthesized policies correctly reflect intended constraints and do not introduce unintended permissions.
This layer contributes to:
\begin{itemize}[nosep]
    \item Normative alignment (R1) by verifying that formalized policies satisfy derived regulatory constraints,

\item Assurance and compliance guarantees (R4) by detecting inconsistencies, conflicts, and unintended authorization paths,

\item Transparency (R6) by generating structured explanations for detected violations or inconsistencies.
\end{itemize}

\paragraph{ Context-Aware Usage Control Enforcement Layer}

(Addresses R2, R3)

The enforcement layer extends traditional access control to satisfy context sensitivity (R2) and lifecycle continuity (R3).
The usage control engine supports:

\begin{itemize}[nosep]
    \item context-aware decisions based on legal basis, purpose limitation, delegation scope, and temporal conditions;
    \item continuous evaluation across pre-access, ongoing, and post-access phases;
    \item enforcement of obligations and re-evaluation as contextual attributes evolve; 
    \item logging of enforcement-relevant state transitions.
\end{itemize}

\paragraph{Runtime Monitoring and Adaptive Feedback Layer}
(Addresses R3, R4, R5, R6)

Even formally verified policies may become misaligned with practice due to evolving organizational contexts or integration errors. Runtime monitoring, therefore, complements static assurance.
This layer:
\begin{itemize}[nosep]
\item observes execution traces and enforcement decisions;
\item detects usage-policy violations;
\item distinguishes policy mis-specification from contextual misuse; and
\item produces structured violation reports.
\end{itemize}

Together, these capabilities provide continuous assurance and support adaptive policy refinement.

\paragraph{The Closed Compliance Loop}

The architectural layers are connected through a continuous compliance loop:

\begin{center}
Regulatory Interpretation $\rightarrow$ Policy Synthesis $\rightarrow$
Formal Assurance $\rightarrow$ Enforcement $\rightarrow$
Monitoring $\rightarrow$ Refinement
\end{center}

Figure 2 summarizes the proposed workflow and its feedback mechanisms.

\section{Conclusion and Future Work}\label{sec:conclusion}
Our analysis indicates that existing access-control approaches are insufficient for next-generation public data exchange infrastructures because they cannot adequately support dynamic, AI-mediated, and cross-organizational data usage. To address these challenges, we identified key requirements for DUC, including normative alignment with regulations, context-aware decision-making, lifecycle control of data usage, assurance mechanisms, adaptability, and transparency with human oversight. Based on these requirements, we proposed a closed-loop framework that connects regulatory interpretation, policy synthesis, formal verification, runtime enforcement, and monitoring into a continuous compliance process.

As future work, we plan to further develop and evaluate the proposed framework through concrete prototypes and case studies in public data exchange infrastructures. In particular, we aim to investigate methods for translating regulatory requirements into formal usage control policies and to explore scalable mechanisms for context-aware enforcement and monitoring in complex cross-organizational data exchange environments. Interdisciplinary research combining usage control, formal verification, requirements engineering, and digital governance is therefore necessary to turn privacy-by-design from a legal principle into an operational property of public data exchange infrastructures.

\bibliographystyle{splncs04}
\bibliography{bibliography}

@String{Springer = "Springer-Verlag" }

@incollection{erree2025justice,
  author = {Peegel, Mari and Turovski, Marcus},
  title = {Justice chancellor: State agencies can access banking secrets without rules},
  year = 2025,
  month = jul,
  day = 2,
  publisher = "ERR.ee"
}

@incollection{erree2025justice2,
  author = {Turovski, Marcus and Tooming, Marko},
  title = {Justice chancellor: 2,000 bank statements viewed without legal basis},
  year = 2025,
  month = jul,
  day = 3,
  published = "ERR.ee"
}

@misc{erree2025entrepreneurs,
  author = {Koppel, Karin and Turovski, Marcus},
  title = {Entrepreneurs take ministry to court over whether their bank accounts were accessed},
  year = 2025,
  month = sep,
  day = 12,
  howpublished  = "ERR.ee"
}

@incollection{erree2025lhv,
  author = {Whyte, Andrew},
  title = {LHV demanding damages of €247 million from FIU over banking data access},
  year = 2025,
  month = jul,
  day = 25,
  howpublished  = "ERR.ee"
}

@incollection{erree2025state,
  author = {Cavegn, Dario},
  title = {State systems illegally passing around personal data on massive scale},
  year = 2017,
  month = apr,
  day = 18,
  howpublished  = "ERR.ee"
}

@incollection{err2025ministry,
  author       = {Karin Koppel and Marcus Turovski},
  title        = {{Ministry not planning to check whether authorities' database queries justified}},
  year         = {2025},
  month        = {October},
  day          = {27},
  howpublished     = {ERR.ee}
}

@online{idsa2025,
  author = {IDSA},
  title = {Interoperability in Data Spaces},
  year = 2025,
  month = jul,
  day = 1,
  urldate = {2025-12-07}
}

@inproceedings{blake2021historical,
  title={A historical analysis on interoperability in Estonian data exchange architecture: perspectives from the past and for the future},
  author={Blake Jackson, Eric and Dreyling, Richard and Pappel, Ingrid},
  booktitle={Proceedings of the 14th ICEGOV},
  pages={111--116},
  year={2021}
}

@article{kalja2002x,
  title={The X-road project},
  author={Kalja, Ahto},
  journal={A project to modernize Estonia’s national databases. Baltic IT\&T review},
  volume={24},
  pages={47--48},
  year={2002}
}

@incollection{xroad2020planetcross,
  author = {Plantera, Federico},
  title = {PlanetCross brings secure data exchange to the Japanese energy sector},
  year = 2020,
  month = jun,
  day = 21,
  publisher = "x-Road.global"
}

@incollection{xroad2024making,
  author = {Petteri, Kivimäki},
  title = {Making of X-Road 8 – PoC June Status Update},
  year = 2024,
  month = jun,
  day = 24,
  publisher = "NIIS.org"
}

@misc{ec2020european,
    author = "{European Commission}",
    title = "A European strategy for data. COM/2020/66 final",
    year = 2020
}

@misc{ec2025data,
    author = "{European Commission}",
    title = "Data Union Strategy. COM(2025) 835 final.",
    year = 2025
}

@article{mcbride2023proactive,
  title={Proactive Public Services--The new standard for digital governments},
  author={McBride, Keegan and Hammerschmid, G and Lume, H and Raieste, Andres},
  pages={1--44},
  year={2023}
}

@techreport{reiberg2022dataspace,
  title        = {What Is a Data Space? Definition of the Concept Data Space},
  author       = {Reiberg, Abel and Niebel, Crispin and Kraemer, Peter},
  type         = {White Paper},
  year         = {2022},
  month        = sep
}

@misc{fraunhofer2026mydata,
  title        = {MYDATA Control Technologies: Informational self-determination through data filtering and masking},
  author       = {{Fraunhofer IESE}},
  year         = {2026},
  url          = {https://www.mydata-control.de/}
}

@inproceedings{novotna2022evaluation,
  title={An evaluation of methodologies for legal formalization},
  author={Novotna, Tereza and Libal, Tomer},
  booktitle={International Workshop on Explainable, Transparent Autonomous Agents and Multi-Agent Systems},
  pages={189--203},
  year={2022},
  organization={Springer}
}

@inproceedings{kosenkov2021vision,
  title={Vision for an artefact-based approach to regulatory requirements engineering},
  author={Kosenkov, Oleksandr and Unterkalmsteiner, Michael and Mendez, Daniel and Fucci, Davide},
  booktitle={Proceedings of the 15th ACM/IEEE International Symposium ESEM},
  pages={1--6},
  year={2021}
}

@inproceedings{talmoudi2024compliance,
  title={Compliance by Design Methodologies in the Legal Governance Schemes of European Data Spaces},
  author={Talmoudi, Kossay and Choukri, Khalid and Gavanon, Isabelle},
  booktitle={Proceedings of the Workshop on Legal and Ethical Issues in Human Language Technologies},
  pages={1--5},
  year={2024}
}

@article{kosenkov2025privacy,
  title={Privacy by design: Aligning GDPR and software engineering specifications with a requirements engineering approach},
  author={Kosenkov, Oleksandr and Zabardast, Ehsan and Fucci, Davide and Mendez, Daniel and Unterkalmsteiner, Michael},
  journal={Information and Software Technology},
  year={2025},
  publisher={Elsevier}
}

@article{puura2025advancing,
  title={Advancing interoperability of data exchange in Europe: Insights from Estonia’s experience for the common European data spaces},
  author={Puura, Anniki and Soe, Ralf-Martin and Thabit, Sara},
  journal={Data in Brief},
  pages={112361},
  year={2025},
  publisher={Elsevier}
}

\end{document}